\documentstyle[pra,twocolumn,aps]{revtex}
   \input epsf
   \begin{document}
   \draft
   \title{Excitation and damping of collective modes of a Bose-Einstein condensate in a one-dimensional lattice}
   \author{O.~Morsch$^1$, J.H.~M\"uller$^1$, D.~Ciampini$^1$, M.~Cristiani$^1$,
   P.B.~Blakie$^2$, C.J.~Williams$^2$, P.S.~Julienne$^2$,
   and E.~Arimondo$^1$}\address{$^1$ INFM, Dipartimento di Fisica E.Fermi, Universit\`{a} di Pisa, Via Buonarroti 2, I-56127 Pisa,Italy \\ $^2$ National Institute of Standards and Technology, Gaithersburg, Maryland 20899-8423.}
   \date{\today}
   \maketitle
   \begin{abstract}
   The mode structure of a Bose-Einstein condensate
   non-adiabatically loaded into a one-dimensional optical lattice
   is studied by analyzing the visibility of the interference
   pattern as well as the radial profile of the condensate after a
   time-of-flight. A simple model is proposed that predicts the
   short-time decrease of the visibility as a function of the
   condensate parameters. In the radial direction, heavily damped
   oscillations are observed, as well as an increase in the
   condensate temperature. These findings are interpreted as a
   re-thermalization due to dissipation of the initial condensate
   excitations into high-lying modes.
   \end{abstract}
   Studying collective modes in a Bose-Einstein condensate (BEC) is an
   efficient method for obtaining information about the dynamics
   of this quantum system~\cite{mewes96,jin96}. So far, both experimental and
   theoretical results have been obtained for low-lying modes of a
   condensate in a magnetic trap, including breathing modes~\cite{chevy02},
   surface modes~\cite{onofrio00} and the scissors mode~\cite{marago00}. Typically,
   in these experiments the collective modes were excited by a
   sudden change in the trap frequency or geometry, and the frequency and
   damping rate~\cite{chikkatur02} of the subsequent oscillations were measured either
   in situ or after a time-of-flight.

   In this paper we present experimental results and some
   preliminary theoretical considerations on the mode structure of
   a Bose-Einstein condensate inside a one-dimensional periodic
   potential. In experiments to date, BECs have been loaded into
   optical lattices mainly in the adiabatic regime in order to
   study, {\em e.g.}, number squeezing~\cite{orzel01} and the Mott-insulator
   transition~\cite{greiner02}. In this context, `adiabatic' refers to the
   modes of the entire condensate rather than the single-well
   oscillation frequency. Therefore, the condition for
   adiabaticity is $\tau_{ramp}>(\mu/\hbar)^{-1}$, where
   $\tau_{ramp}$ is the time over which the periodic
   potential is ramped up, and $\mu$ is the chemical potential of
   the condensate~\cite{band02}. By violating this condition (but still
   satisfying $\tau_{ramp}>2\pi/\omega_{lat}$, with $\omega_{lat}$
   the single-well harmonic oscillator frequency), collective modes are excited in the condensate.

   Our experimental setup is described in detail
   elsewhere~\cite{morsch01,cristiani02}. After creating BECs of $N_0=1.5(5)\times 10^4$
   $^{87}$Rb atoms\cite{footnote_uncertain} in a triaxial time-orbiting potential trap, we
   adiabatically lower the trap frequency $\overline{\nu}_{trap}$ to the desired value and
   then superimpose onto the magnetic trap an optical lattice
   along the vertical trap axis (for which the trap frequency
   $\nu_{long}\approx\overline{\nu}_{trap}$)
   created by two linearly polarized Gaussian laser beams
   intersecting at a half-angle $\theta=18$ degrees and detuned by $\approx 30\,\mathrm{GHz}$ above the
   rubidium resonance line. The periodic
   potential $V(z)=V_0\sin^2(\pi z/d)$ thus created has a lattice spacing
   $d=1.2\,\mathrm{\mu m}$, and the depth $V_0$ of the potential
   (measured in lattice recoil energies $E_{rec}=\hbar^2
   \pi^2/2md^2$) can be varied between $0$ and $\approx
   20\,E_{rec}$ by adjusting the laser intensity using
   acousto-optic modulators. The number of lattice sites occupied
   by the condensate lay between $10$ and $15$, depending on the
   trap frequency.

   \begin{figure}
   \centering\begin{center}\mbox{\epsfxsize 2.65 in \epsfbox{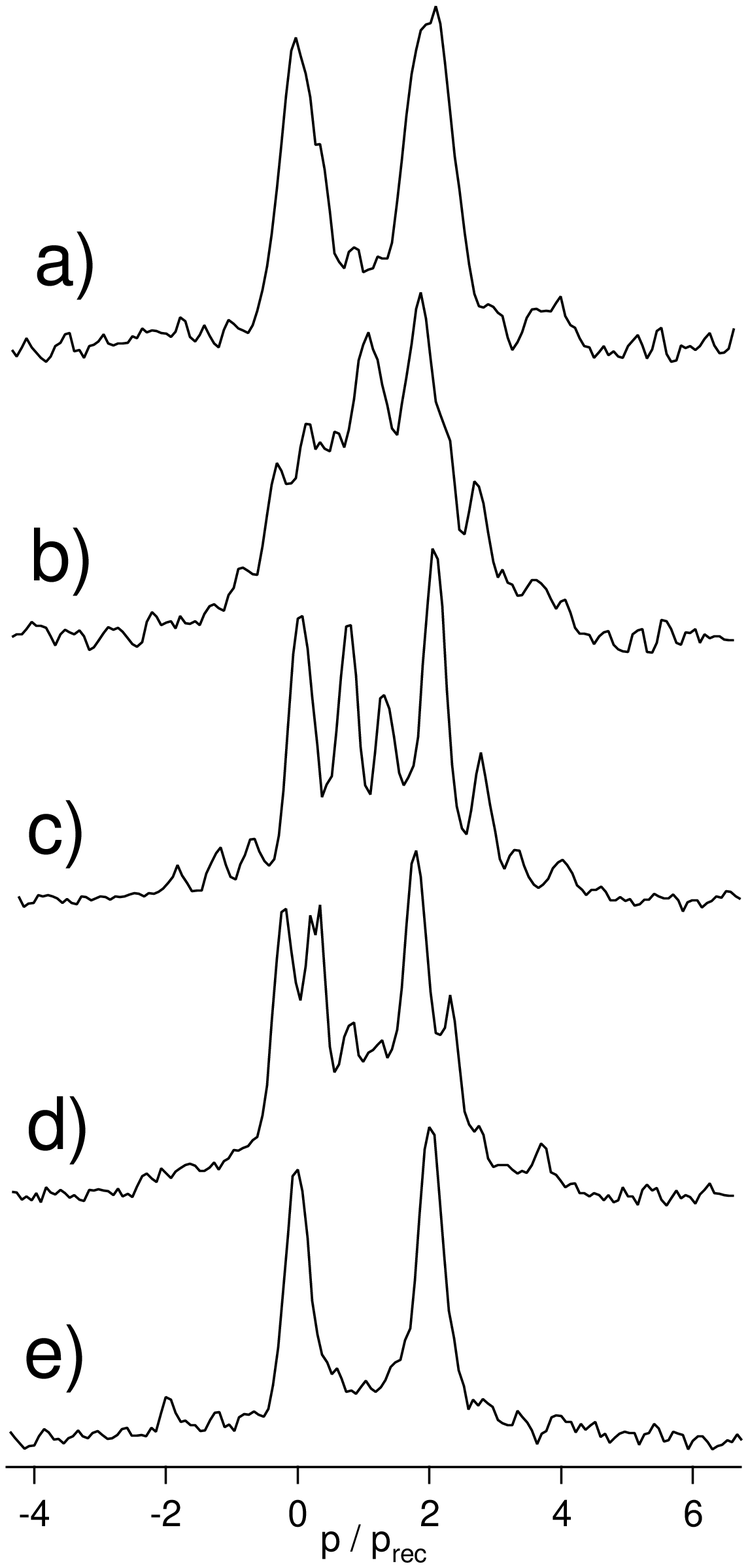}}
   \caption{Evolution of the interference pattern (integrated perpendicular to the lattice direction) of a condensate released from an optical lattice after
   non-adiabatic loading. The distinct two-peaked structure visible immediately after loading (a) is washed out within the first few milliseconds
   (b), after which the interference pattern takes on a complex
   structure ((c) and (d)). For long holding times, the initial two-peaked structure
   reappears (e). In this experiment, $\overline{\nu}_{trap}=26.7\,\mathrm{Hz}$,
   $\tau_{ramp}=5\,\mathrm{ms}$, and $V_0\approx 15\,E_{rec}$. From (a) to (e),
   $t_{hold}=1,22,50,100$ and $300\,\mathrm{ms}$, respectively.}\label{Fig_1}
   \end{center}\end{figure}

   In a typical experiment, the optical lattice was ramped up in
   $\tau_{ramp}\approx 1-5 \,\mathrm{ms}$, after which the
   potential was kept at its maximum value $V_0$ for a holding time
   $t_{hold}$. At the end of the holding time, the lattice was accelerated in $1\,\mathrm{ms}$ to one (lattice) recoil velocity by
   chirping the frequency difference between the lattice beams. Immediately after that,
   both the lattice and the magnetic trap were switched
   off.  After a time-of-flight of $20-22\,\mathrm{ms}$, the
   expanded condensate was imaged using a resonant probe flash.
   Figure~\ref{Fig_1} shows typical integrated absorption images obtained in
   this manner for different holding times. For short
   times, a clean double-peak structure is visible, as expected
   from the interference between the condensates expanding from
   the individual lattice wells (with a $\pi$ phase difference
   between them due to the final acceleration). During the first
   few milliseconds, this pattern evolves into a more complicated
   structure featuring several additional peaks, and finally
   washes out completely, resulting in a single
   Gaussian-shaped lump. In an intermediate regime ($t_{hold}\approx 20-100\,\mathrm{ms}$), the interference pattern again becomes complex, with
   multiple peaks. For longer waiting times, the two-peaked
   structure reappears.

   We analyzed our experimental data in two different ways. In the
   lattice direction, we characterized the interference pattern (integrated perpendicular to the lattice direction)
    through
   a visibility $\xi$ calculated as
   \begin{equation}
  \xi=\frac{h_{peak}-h_{middle}}{h_{peak}+h_{middle}},
   \end{equation}
   where $h_{peak}$ is the mean value of the absorption image at
   the position of the two peaks, and $h_{middle}$ is the value of
   the absorption image mid-way between the two peaks (averaged
   over a range of $1/5$ of the peak separation). In the
   direction perpendicular to the optical lattice, we fitted a bimodal Gaussian function to
   the (longitudinally) integrated absorption profile and
   extracted from this fit the condensate fraction and the width $\rho_{perp}$ of the condensate part.

   \begin{figure}
   \centering\begin{center}\mbox{\epsfxsize 2.8 in \epsfbox{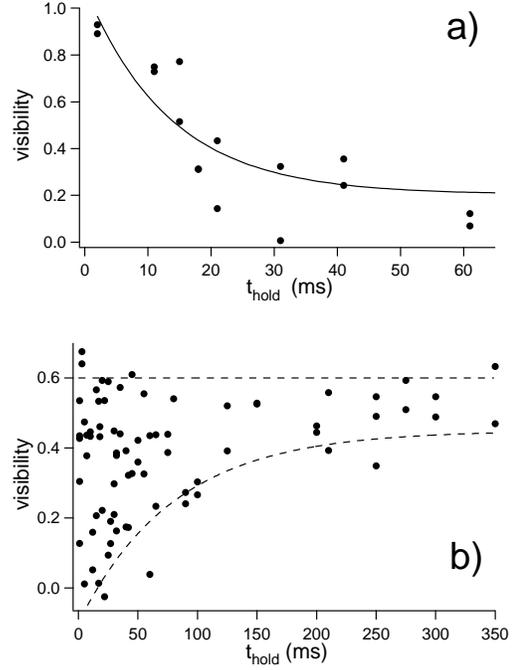}}
   \caption{Examples of the short-term (a) and long-term behaviour (b) of the visibility for a condensate in an optical lattice after non-adiabatic
   loading. In (a), the initial loss of visibility is evident (solid line: exponential fit with time constant
   $13\,\mathrm{ms}$). The trap frequency
   $\overline{\nu}_{trap}=19.8\,\mathrm{Hz}$, and
   $V_0/E_{rec}\approx 15$ with $\tau_{ramp}=1\,\mathrm{ms}$.
   In (b), after strong initial fluctuations the visibility approaches a stable value close to
   the initial visibility. In this experiment, $\overline{\nu}_{trap}=26.7\,\mathrm{Hz}$, $V_0/E_{rec}\approx 15$ and $\tau_{ramp}=5\,\mathrm{ms}$.
   The dashed lines are an envelope to the data points with a constant upper part and an exponential lower part with
   a time constant of $\approx 80\,\mathrm{ms}$.}\label{Fig_2}
   \end{center}\end{figure}

   Figure~\ref{Fig_2} shows examples of the short-term and
   long-term behaviour of $\xi$. Initially, $\xi$ rapidly decreases
   from a value of $\approx 0.6-0.9$ to roughly $0$ within
   $\approx 5-20\,\mathrm{ms}$, depending on the trap frequency
   $\overline{\nu}_{trap}$. Subsequently, $\xi$ typically rises
   again and begins to fluctuate in an apparently random manner
   for a few tens of milliseconds. Finally, these fluctuations die
   out and $\xi$ stabilizes at a value close to the initial
   visibility.

\begin{figure}
   \centering\begin{center}\mbox{\epsfxsize 2.8 in \epsfbox{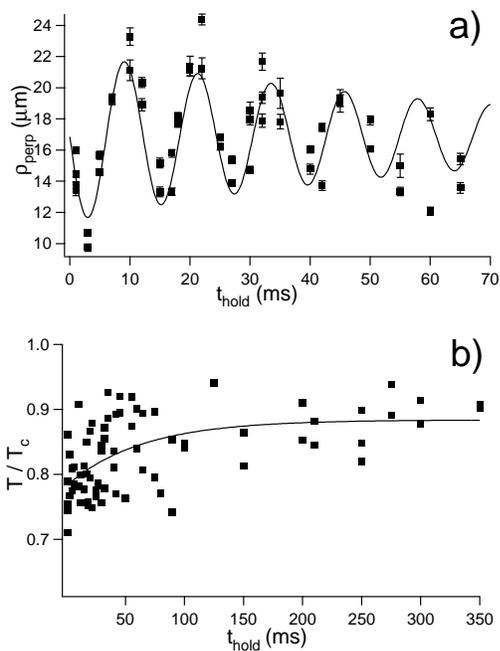}}
   \caption{Time evolution of the width (a) and the temperature (b) of the condensate in the optical lattice (same trap and lattice
   parameters as in Fig.~\ref{Fig_2} (b)).
   The solid lines are are fits to the data using an exponentially damped sinusoidal oscillation in (a) and
   a simple exponential function in (b). In both cases, the time constants of the exponential part are
   $\approx 70\,\mathrm{ms}$. Note the different time scales in the two graphs.}\label{Fig_3}
   \end{center}\end{figure}

   The corresponding behaviour of the condensate width $\rho_{perp}$ and
   temperature (calculated from the condensate fraction) is shown
   in Fig.~\ref{Fig_3}. One clearly sees radial oscillations of
   $\rho_{perp}$ at a frequency $\nu_{osc}=2\nu_{perp}$,
   where $\nu_{perp}=\sqrt{2}\overline{\nu}_{trap}$ is the trap frequency
   perpendicular to the lattice direction.
   These oscillations are heavily damped with a quality factor
   $Q=2\pi\nu_{osc}/\gamma_{damp}\approx 10-30$. Simultaneously,
   the condensate temperature expressed as $T/T_c$ (where $T_c$ is
   the critical temperature for the BEC transition) increases and
   approaches a new steady-state value on a time scale comparable
   to $1/\gamma_{damp}$.

   The short-term decrease in $\xi$ can be explained using a simple model.
   We take the Bose-Einstein condensate wavefunction to evolve
according the Gross-Pitaevskii equation
\begin{eqnarray}
i\hbar\frac{\partial}{\partial t}\psi
&=&\Big[-\frac{\hbar^2}{2m}\nabla^2+\frac{1}{2}m\sum_j
\omega_j^2x_j^2\nonumber\\ &&+ V_0(t)\sin^2\left(\frac{\pi
z}{d}\right)+U|\psi|^2\Big]\psi,\label{tdGPE}
\end{eqnarray}
where $x_j=\{x,y,z\}$ are the spatial coordinates along which the
respective trap frequencies are $\omega_j$ ($=2\pi\nu_j$), $U=4\pi
a\hbar^2/m$ is the collisional interaction strength with
$a\approx5.4\,\mathrm{nm}$ the $s$-wave scattering length.

The rate at which the lattice is raised is slow enough that band
excitations can be ignored and the condensate density will be
reshaped to lie at the potential minima of each lattice site.
Because we are interested in the phase properties of the entire
condensate, it is convenient to consider a continuous Wannier
\cite{ZimanBook} or envelope representation \cite{Steel1998a} of
the wavefunction, where we define the envelope $f(\mathbf{x},t)$
to represent the slowly varying amplitude of the Gross-Pitaevskii
wavefunction $\psi(\mathbf{x},t)$, in which the rapid density
variation along the lattice is smoothed out (also see
\cite{kramer2002a}). The evolution equation for $f$ is of a
similar form to Eq. (\ref{tdGPE}) except that the lattice
potential no longer appears explicitly; the diffusion along $z$ is
generated by the Bloch dispersion relation reflecting the modified
tunneling properties in this direction; and the mean-field term is
renormalized as $U|\psi|^2\to(1+\tilde{c})U|f|^2$ resulting from
the compressed condensate density in the lattice. In the tight
binding limit the Wannier states are localized and can be
approximated by the harmonic oscillator orbital
$w(z)=\exp(-z^2/2a_{ho}^2)/\sqrt[4]{\pi a_{ho}^2}$, where
$a_{ho}=\sqrt{\hbar/m\omega_{lat}}$ is the oscillator length and
$\omega_{lat}=\pi\sqrt{2V_0/md^2}$ is the oscillation frequency
about the lattice minima.
 In this limit the renormalized mean-field term is given by
 $\tilde{c}=(d\int |w(z)|^4dz-1) =(\sqrt[4]{mV_0d^2/2\hbar^2}-1)$
 \cite{constnote}, with the validity condition $a\ll a_{ho}\ll d$.

To develop an approximate expression for the short time dephasing
behavior of the condensate we assume that during the lattice
loading and subsequent time over which the dephasing occurs
density transport in the condensate is negligible, i.e. the
evolution occurs in the phase of the wavefunction. In the lattice
direction the tunneling frequency determines the time scale over
which this assumption will be valid, which is typically of order
$100\,\mathrm{ms}$ in experiments where dephasing is observed. We
also find that for a given trap frequency, there always exists a
minimum lattice depth below which no dephasing occurs.

Expressing the envelope function in terms of amplitude and phase
as $f=|f|\exp(iS)$, and neglecting the effects of spatial diffusion, the
envelope equation of motion reduces to
\begin{eqnarray}
-\hbar\frac{\partial}{\partial t} S({\mathbf{x}},t) &=&
\frac{1}{2} m\sum_j\omega_j^2x_j^2 +
U(1+\tilde{c})|f|^2.\label{phaseEvolve}
\end{eqnarray}
Ignoring density transport $|f|^2$ can be approximated by the
Thomas-Fermi density profile for the condensate in the initial
harmonic trap, which is of the form
$|\psi_{\rm{TF}}(\mathbf{x})|^2=n_0[1-\sum_j({x_j}/{R_j})^2]$,
where $n_0$ is the peak density and $R_j=\{R_x,R_y,R_z\}$ are the
Thomas-Fermi radii \cite{Dalfovo1999a}. With this substitution the
phase evolves as
\begin{equation} S({\mathbf{x}},t)=-\left(\frac{\tilde{c}n_0 U
t}{\hbar}\right)\left(\frac{z}{R_z}\right)^2+S_0(t),\label{phase}
\end{equation}
where $S_0(t)$ is a spatially constant phase. This result shows
that the condensate develops a quadratic phase profile along the
lattice arising from the imbalance of meanfield and harmonic trap
potential energy in the lattice, and is equivalent to a spread in
the quasi momentum distribution of the system.
\begin{figure}
\centering\begin{center}\mbox{\epsfxsize 2  in
\epsfbox{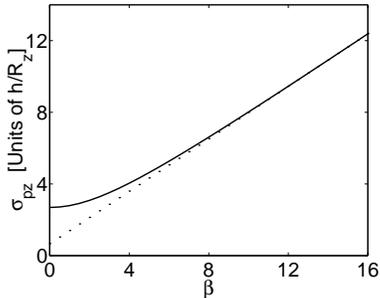}} \caption{\label{figBeta} Momentum width of a
condensate with a Thomas-Fermi density distribution and a
quadratic phase profile. Numerical calculation (solid), linear fit
(dotted). }\end{center}
\end{figure}
For a large lattice the quasi momentum distribution of the
condensate and the momentum distribution of the envelope function
are identical if the momentum distribution lies entirely within
the first Brillouin zone. We have calculated the rms-width of the
$z$-component of momentum  $ \sigma_{pz}$  for a wavefunction of
the form $\psi(\mathbf{x},\beta)=\psi_{\rm{TF}}(\mathbf{x})
\exp\left(-i\beta\left(z/R_z\right)^2\right)$, where $\beta$ is
the total phase difference between the center and the outside of
edge of the condensate. The results (shown in Fig. \ref{figBeta})
indicate that for $\beta\gtrsim2$, the momentum width linearly
increases with $\beta$ and is well approximated by $\sigma_{pz}
(\beta)\approx 0.73\,\hbar\beta/R_z$. Identifying $\beta$ with the
coefficient of the quadratic spatial phase term in Eq.
(\ref{phase}), we see that the width of the momentum distribution
increases linearly with time.

\begin{figure}
   \centering\begin{center}\mbox{\epsfxsize 2.4 in \epsfbox{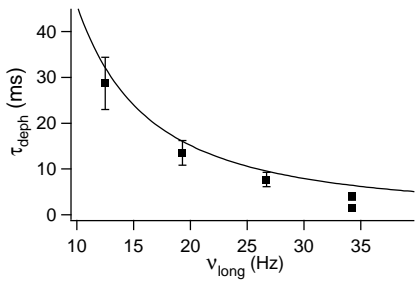}}
   \caption{Experimental dephasing times as a function of the trap frequency $\nu_{long}$ in the lattice
   direction. The solid line is the theoretical prediction for $\tau_{deph}$ (see text).}\label{Fig_5}
   \end{center}\end{figure}

The condensate will initially appear dephased when the quasi
momentum width significantly fills the first Brillouin zone, which
has a half width of $h/2d$. The exact portion of the Brillouin
zone which must be filled depends on the observable used to
determine the dephasing and may be taken as a fitting parameter.
Here we take
 $\sigma_{pz}=h/4d$ as the requirement for dephasing, which
can be inverted using expression (\ref{phase}) to yield the
dephasing time
 \begin{equation}
\tau_{deph} = \frac{ R_z h}{2.9\,\tilde{c} d n_0 U}.
\end{equation}
Figure~\ref{Fig_5} shows $\tau_{deph}$ as a function of
$\nu_{long}$ (which determines $R_z$ and $n_0$) together with
experimentally measured dephasing times. Equally good agreement
with experiment is found when calculating $\tau_{deph}$ for the
parameters of~\cite{orzel01}.

   For the intermediate and long-term behaviour of the visibility, the
   phase-winding model predicts partial and complete revivals of the relative phases,
   leading to a complicated time-evolution of $\xi$. Experimentally,
   we see a rather erratic behaviour of $\xi$ for intermediate
   times, with the visibility fluctuation between $0-0.2$ and $0.5-
   0.6$, followed by a stabilization at a high value around $0.6$.
   The fact that this stabilization takes place on the same
   time-scale as the damping of the radial oscillations and the
   increase in temperature of the condensate leads us to speculate
   that all these phenomena are related through a dissipation
   mechanism whereby the the initial excitations in the
   longitudinal direction ({\em i.e.} reflected in the phase
   differences between adjacent lattice sites) and in the radial
   directions lead to a re-thermalization of the system at a
   higher temperature and hence lower condensate fraction, with
   the remaining condensate now in the ground state of the
   combined harmonic and periodic trapping potentials. More
   detailed theoretical investigations in this direction are
   planned for the future.

   We also note here that the tunneling
   time between adjacent wells is also comparable to the damping time
   of the visibility fluctuations for the parameters of our
   experiment. In order to understand better the importance of the
   various mechanisms, i.e. damping with a possible coupling
   between the different modes, and re-phasing due to tunneling,
   it will be important in future to conduct experiments for a
   range of different parameter combinations for the trap and the lattice. Also, repeating our
   experiment in a configuration in which more lattice sites are
   occupied by the condensate will most likely eliminate the
   intermediate revivals (which require a large fraction of the
   sites to be in phase, which is less likely for a large number
   of sites) and leave only the lower envelope (see
   Fig.~\ref{Fig_2}) of the visibility evolution. Finally, the
   role of the finite temperature ($T\approx 0.7\,T_c$) at which we start our experiment
   will have to be investigated more closely.

   In summary, we have studied the behaviour of a Bose-Einstein
   condensate non-adiabatically loaded into a one-dimensional
   optical lattice. The time-evolution of the collective modes
   thus excited has been characterized through the visibility of
   the interference pattern as well as its radial width. Both
   quantities exhibit large initial variations that are strongly
   damped on a time-scale comparable to the relaxation time of the
   condensate temperature.

   This work was supported by the MURST (PRIN2000
   Initiative), the INFM (Progetto di Ricerca Avanzata
   `Photonmatter'), and by  the EU through the Cold Quantum
   Gases Network, Contract No. HPRN-CT-2000-00125. P.B.B., C.J.W.,
   and P.S.J. acknowledge support from
   the U.S. Office of Naval Research. O.M. gratefully
   acknowledges financial support from the EU within the IHP
   Programme.


\begin{references}
   \bibitem{mewes96} M.O.~Mewes, M.R.~Andrews, N.J.~van Druten,
   D.M.~Kurn, D.S.~Durfee, C.G.~Townsend, and W.~Ketterle, Phys.
   Rev. Lett. {\bf 77}, 988 (1996).
   \bibitem{jin96} D.S.~Jin, M.R.~Matthews, J.R.~Ensher,
   C.E.~Wieman, and E.A.~Cornell, Phys. Rev. Lett. {\bf 78}, 764
   (1996).
   \bibitem{chevy02} F.~Chevy, V.~Bretin, P.~Rosenbusch,
   K.W.~Madison, and J.~Dalibard, Phys. Rev. Lett. {\bf 88},
   250402 (2002).
   \bibitem{onofrio00} R.~Onofrio, D.S.~Durfee, C.~Raman, M.~Kohl,
   C.E.~Kuklewicz, and W.~Ketterle, Phys. Rev. Lett. {\bf 84}, 810
   (2000).
   \bibitem{marago00} O.M.~Marag\`o, S.A.~Hopkins, J.~Arlt,
   E.~Hodby, G.~Hechenblaikner, and C.J.~Foot, Phys. Rev. Lett.
   {\bf 84}, 2056 (2000).
   \bibitem{chikkatur02}Damping of condensate excitations due to
   condensate merging was recently observed by A.P.~Chikkatur {\em et
   al.}, Science {\bf 296}, 2193 (2002).
   \bibitem{orzel01} C.~Orzel, A.K.~Tuchman, M.L.~Fenselau,
   M.~Yasuda, and M.A.~Kasevich, Science {\bf 291}, 5512 (2001).
   \bibitem{greiner02} M.~Greiner, O.~Mandel, T.~Esslinger,
   T.W.~H\"ansch, and I.~Bloch, Nature (London) {\bf 415}, 6867
   (2002).
   \bibitem{band02} Y.B.~Band and M.~Trippenbach, Phys. Rev. A {\bf
   65}, 053602 (2002).
   \bibitem{morsch01} O.~Morsch, J.H.~M\"uller, M.~Cristiani,
   D.~Ciampini, and E.~Arimondo, Phys. Rev. Lett. {\bf 87}, 140402
   (2001).
   \bibitem{cristiani02}M.~Cristiani, O.~Morsch, J.H.~M\"uller,
   D.~Ciampini, and E.~Arimondo, Phys. Rev. A {\bf 65}, 063612 (2002).
   \bibitem{footnote_uncertain}All uncertainties reported here are one standard deviation combined systematic and statistical
uncertainties.
   \bibitem{kramer2002a} M. Kr\"amer, L. Pitaevskii, and S. Stringari,
   Phys. Rev. Lett. {\bf 88}, 180404 (2002).
   \bibitem{ZimanBook} J. M. Ziman, {\em Principles of the Theory
   of Solids}, Cambridge University Press (1964).
   \bibitem{Steel1998a} M. J. Steel and Weiping Zhang, cond-mat/9810284
   (1998).
 \bibitem{Dalfovo1999a}F. Dalfovo, S. Giorgini, L. P. Pitaevskii, and S. Stringari,
 Rev. Mod. Phys.  {\bf 71}, 463-512 (1999)

\bibitem{constnote} We note that  $\tilde{c}$ relates to the
renormalized coupling strength $\tilde{g}$ defined in
\cite{kramer2002a} according to $\tilde{g}=U(1+\tilde{c})$.
    \end{references}
   \end{document}